\documentclass[11pt]{article}
\usepackage{moriond,epsfig}

\def\be{\begin{equation}}
\def\ee{\end{equation}}
\def\bea{\begin{eqnarray}}
\def\eea{\end{eqnarray}}

\usepackage{graphicx,psfrag}
\begin{document}
\vspace*{4cm}
\title{CALCULATION OF QCD LOOPS USING TREE-LEVEL MATRIX ELEMENTS}

\author{JAN WINTER \footnote{%
    Talk presented at the 45th Rencontres de Moriond, Workshop on QCD
    and High Energy Interactions, March 13-20, 2010 at La Thuile, Italy.}}

\address{Theoretical Physics Department,\\
  Fermi National Accelerator Laboratory, P.O.\ Box 500, Batavia, IL 60510, USA}

\maketitle\abstracts{The possibility of treating colour in one-loop
  amplitude calculations alike the other quantum numbers is briefly
  discussed for semi-numerical algorithms based on generalized
  unitarity and parametric integration techniques. Numerical results
  are presented for the calculation of virtual corrections in
  multi-gluon scattering.}

\begin{flushright}
  \vspace*{-72mm}
  {\small FERMILAB-CONF-10-118-T}\\[67mm]
\end{flushright}

\section{An algorithmic solution based on generalized unitarity for
  the evaluation of colour-dressed one-loop amplitudes}
\label{sec:one}

The evaluation of QCD higher-order corrections in particular for
multi-particle processes is indispensable for a good understanding of
total and differential cross sections at hadron colliders such as the
Tevatron or the LHC. The automation of these calculations at the
next-to-leading order (NLO) in the strong coupling constant has been a
very appealing goal ever since the Monte Carlo tools dealing with the
computation of tree-level cross sections have matured and been greatly
optimized. Over the past few years a number of subtraction codes has
been developed handling the cancellations of singularities occurring
in NLO calculations in a general way. Accordingly a standard interface
has lately been agreed on between Monte Carlo tools and one-loop
matrix-element programs~\cite{Binoth:2010xt,Binoth:2010ra}. Recently
these programs have also shown a remarkable advancement. New
techniques based on combining generalized unitarity and parametric
integration methods have become available and made the computation of
multi-leg one-loop corrections feasible. Prominent examples are given
by the vector boson plus 3 jet NLO predictions provided by the
{\sc BlackHat} and {\sc Rocket}
groups~\cite{Berger:2009ep,Berger:2010vm,KeithEllis:2009bu,Melnikov:2009wh}
and the $t\bar t$\/ plus 1 jet NLO result~\cite{Melnikov:2010iu}.
These calculations separate the treatment of the colour quantum
numbers from the other degrees of freedom, which makes it harder to
fully automatize these approaches. In a recent
publication~\cite{Giele:2009ui} it has been shown that the
Ellis--Giele--Kunszt--Melnikov algorithm~\cite{Ellis:2007brGiele:2008ve}
can be extended to treat colour along the same line with the other
quantum numbers. Given the implementation for ordered gluon one-loop
amplitudes~\cite{Winter:2009kd}, the important changes are: the sums
over ordered cuts present in the decomposition of the one-loop
integrands and amplitudes are replaced by sums over unordered
partitions including their non-cyclic and non-reflective permutations.
Gluon bubble contributions now come with a symmetry factor of $1/2!$.
The integrand's residues are calculated through products of
colour-dressed tree-level amplitudes obtained from dressed recursion
relations where internal colour degrees of freedom are summed over.
Finally, to extract the lower-point coefficients, the subtraction
terms due to higher-cut contributions are identified by a de-pinching
procedure that may involve loop-momentum shifts by external momenta.

\section{Numerical calculation of multi-gluon one-loop corrections}
\label{sec:two}

Taking the example of multi-gluon scattering, the performance and
major results of the algorithm briefly introduced in
Section~\ref{sec:one} are discussed below. A more comprehensive
description of the outcomes of these numerical calculations can be
found in the original paper~\cite{Giele:2009ui}.

\begin{figure}[t!]
\centerline{
  \psfig{figure=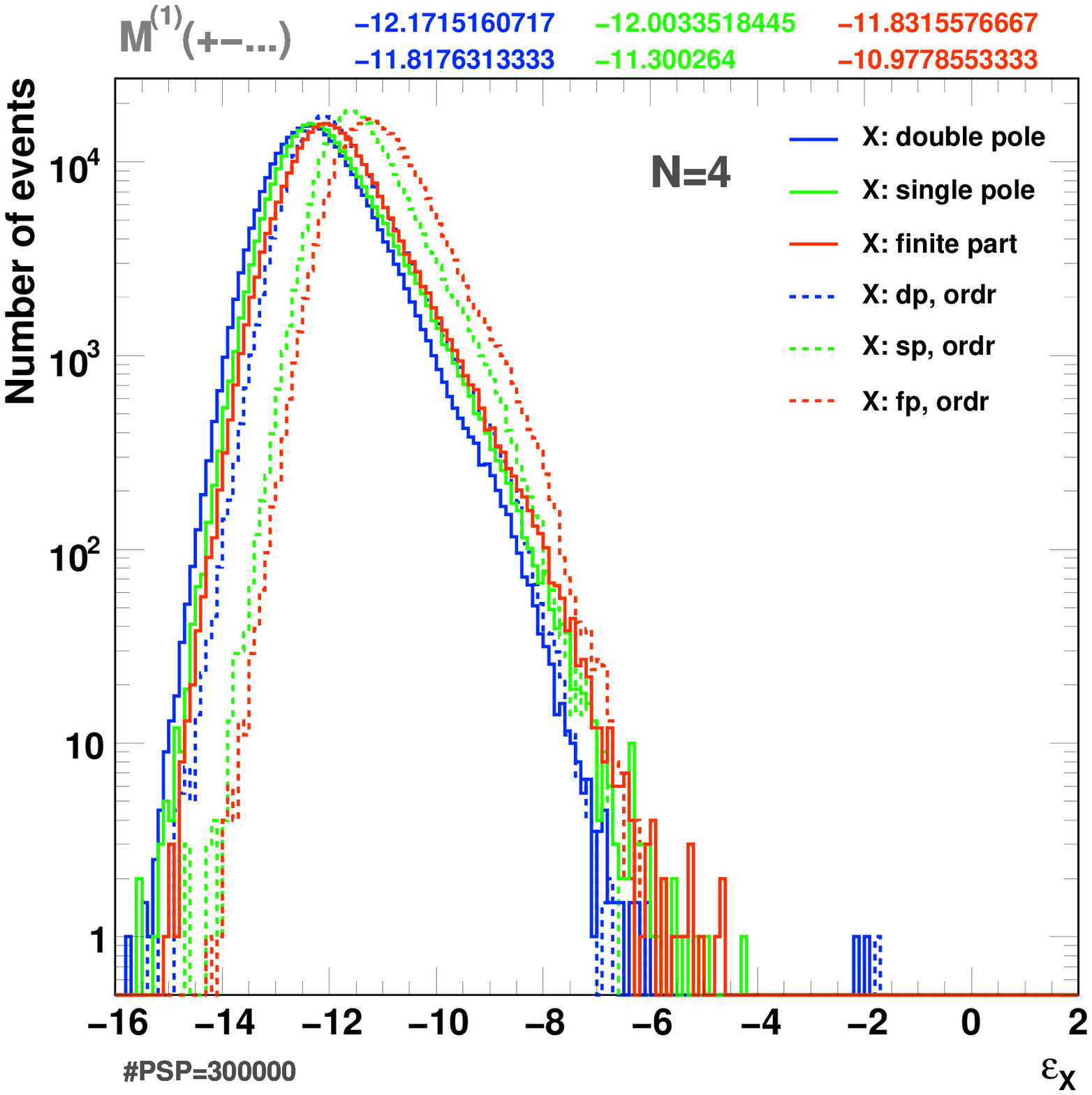,width=55mm,angle=0}\hskip11mm
  \psfig{figure=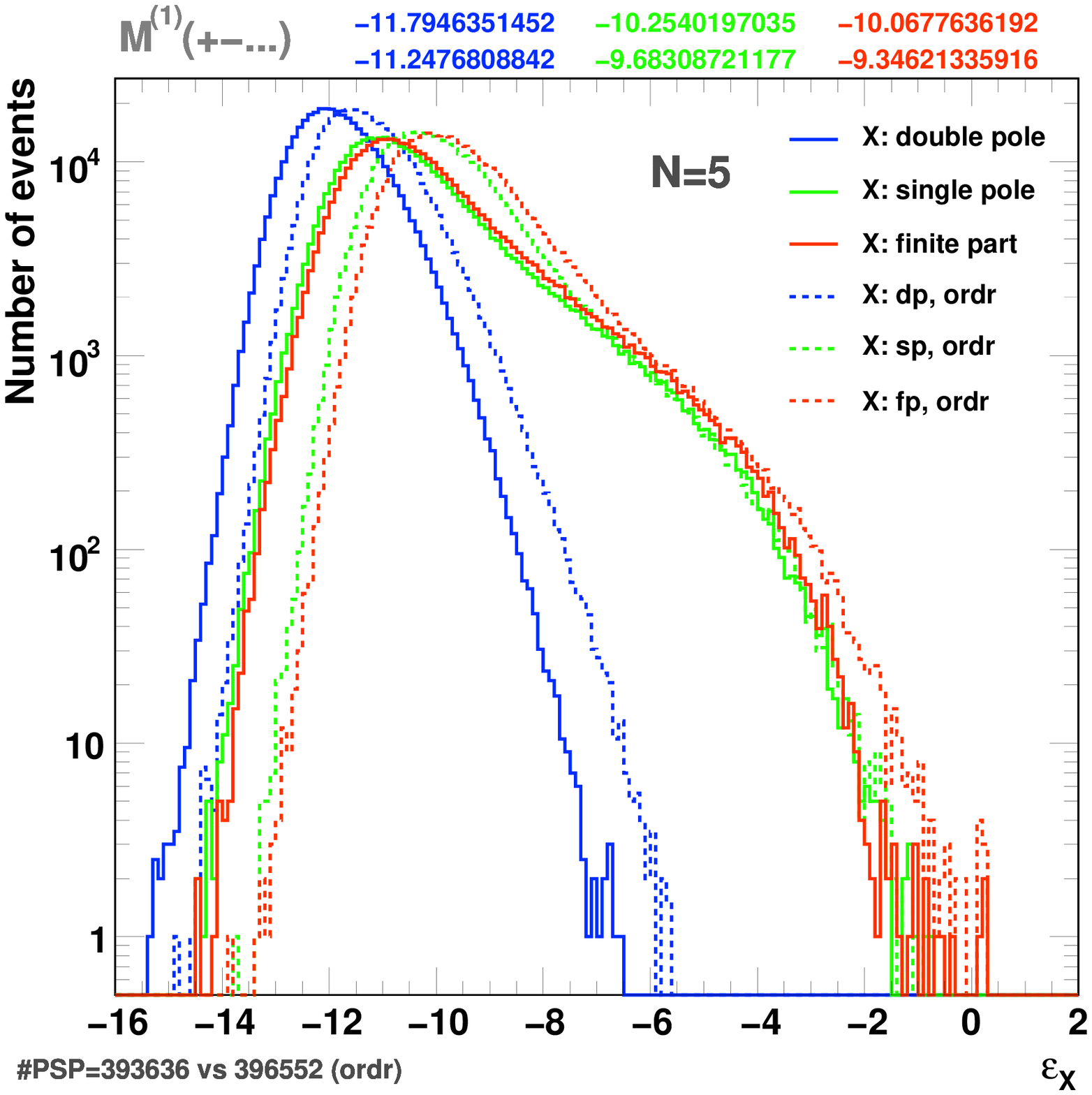,width=55mm,angle=0}}
\centerline{
  \psfig{figure=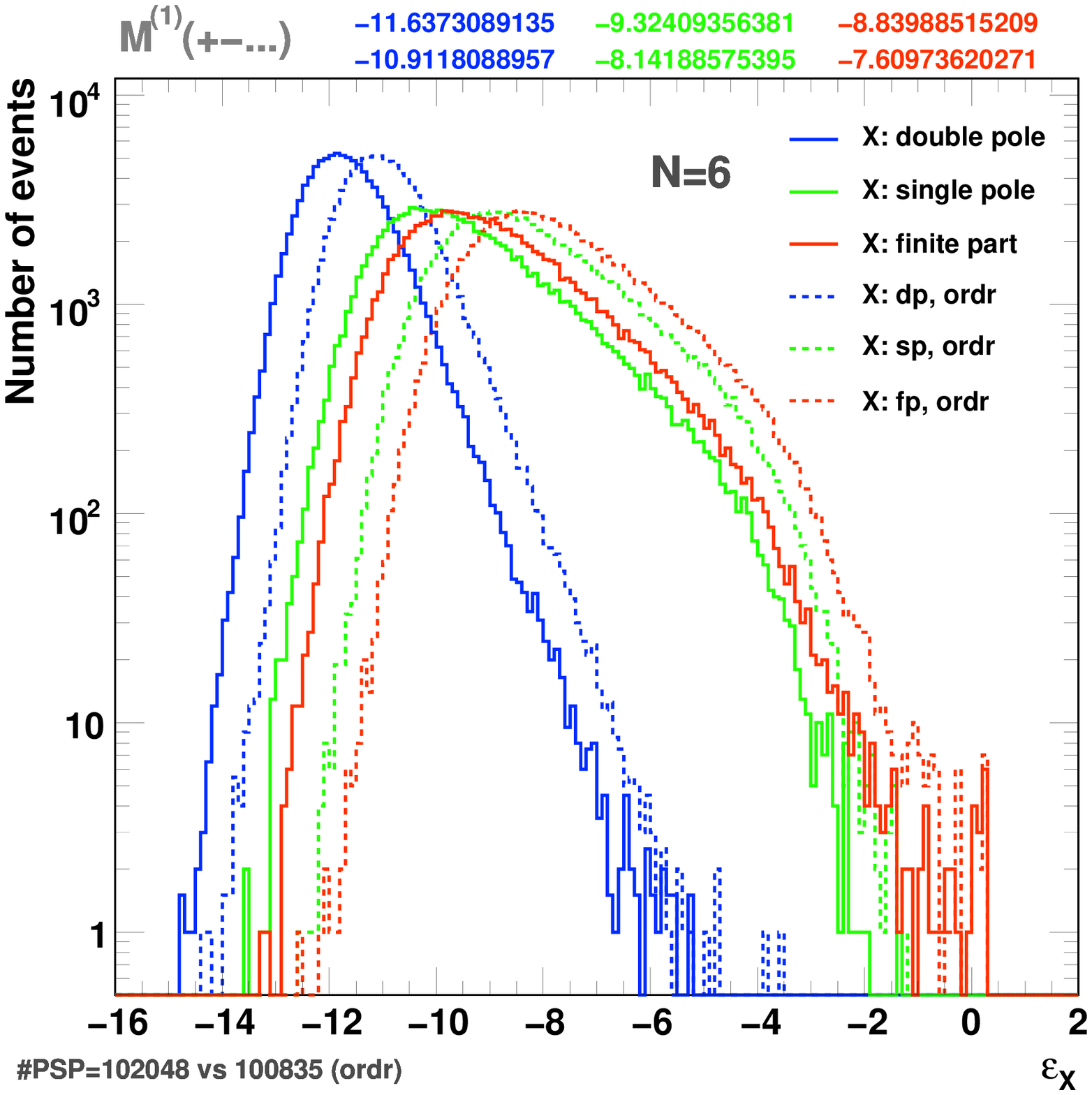,width=55mm,angle=0}\hskip11mm
  \psfig{figure=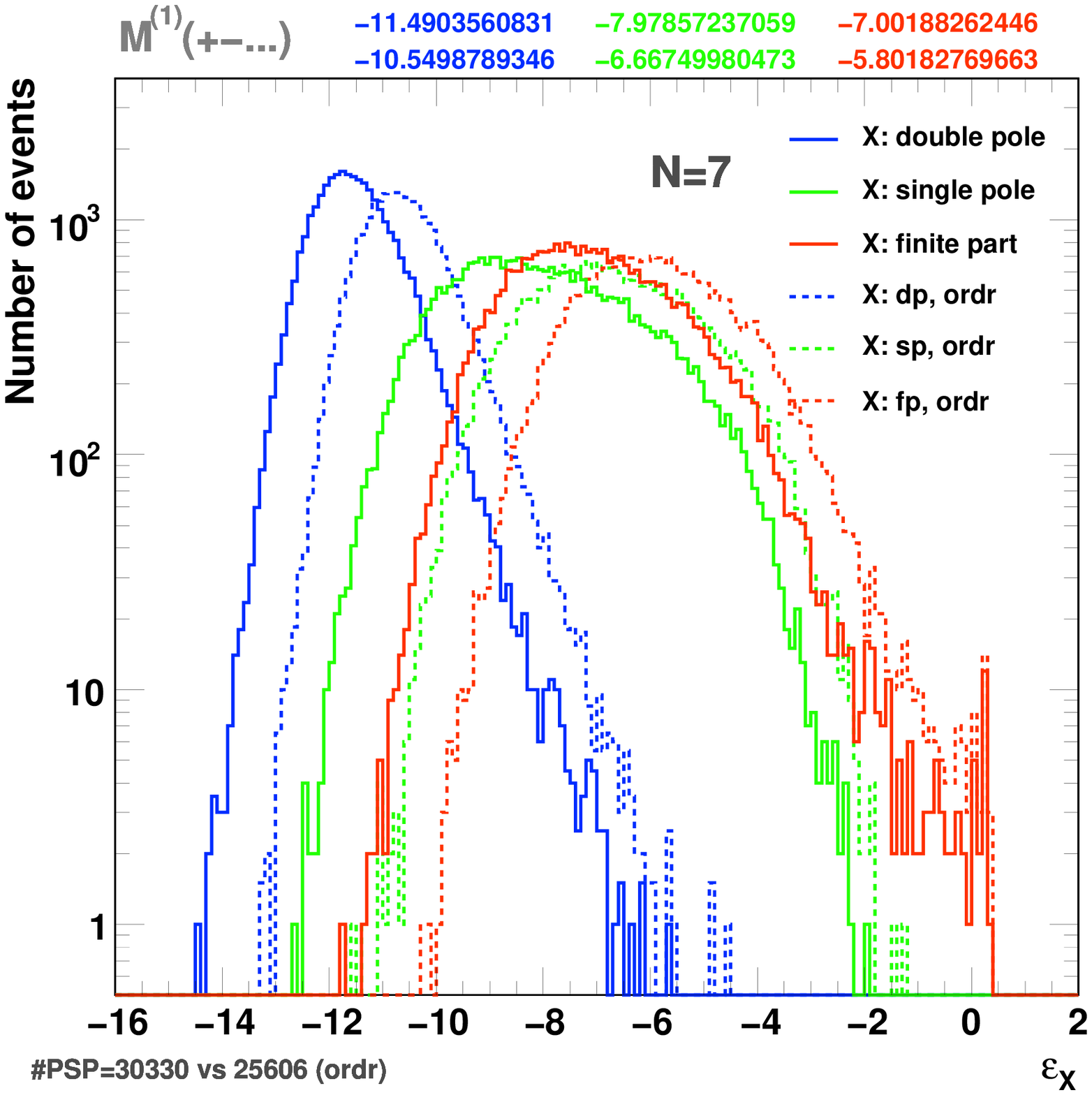,width=55mm,angle=0}}
\vskip-2mm
\caption{Double-, single-pole and finite-part accuracy distributions
  obtained from double-precision computations of one-loop amplitudes
  for $N=4,5,6,7$ gluons with polarizations $\lambda_k=+-+-\ldots$ and
  randomly chosen non-vanishing colour configurations. Results from
  the colour-dressed algorithm are compared to those of the
  colour-ordered method (labeled ``ordr'') indicated by dashed
  curves. The mean accuracies and the number of randomly picked
  phase-space points (subject to the cuts detailed in the text) are
  displayed at the top and bottom left of the plots, respectively.
  Phase-space points required to be calculated at higher precision
  were vetoed.
  \label{fig:accus}}
\end{figure}
The convenient way to cross-check the results of the colour-dressed
algorithm for the full one-loop amplitudes for $N$\/ gluons is to make
use of the colour decomposition approach. Given the polarization
states $\lambda_k\in\{+,-\}$ of the $k=1,\ldots,N$\/ gluons and their
colours $(ij)_k$ using the colour-flow notation $i_k,j_k\in\{1,2,3\}$,
one sums up all relevant ordered amplitudes after having multiplied
them by their corresponding colour factors. The colour-ordered
$N$-gluon one-loop amplitudes are computed with the
algorithm~\cite{Winter:2009kd} based on the original
Ellis--Giele--Kunszt--Melnikov method.
The quality of the one-loop amplitude determination can be analyzed by
means of the logarithmic relative deviations of the double (dp) and
single poles (sp) and the finite part (fp). They are defined as
follows:
\be
\varepsilon_{\rm dp}\;=\;
\log_{10}\,\frac{|{\cal M}^{(1)[1]}_{\rm dp,num}-{\cal M}^{(1)}_{\rm dp,th}|}
                {|{\cal M}^{(1)}_{\rm dp,th}|}\,,\qquad
\varepsilon_{\rm s/fp}\;=\;
\log_{10}\,\frac{2\,|{\cal M}^{(1)[1]}_{\rm s/fp,num}-
                     {\cal M}^{(1)[2]}_{\rm s/fp,num}|}
                {|{\cal M}^{(1)[1]}_{\rm s/fp,num}|+
                 |{\cal M}^{(1)[2]}_{\rm s/fp,num}|}\ ,
\ee
Two independent solutions denoted by $[1]$ and $[2]$ are used to test
the accuracies of the single poles and finite parts. The analytic
structure of the double poles is particularly simple and proportional
to the Born amplitudes, ${\cal M}^{(1)}_{\rm dp,th}=
-(c_\Gamma/\epsilon^2)\,N\,N_{\rm C}\,{\cal M}^{(0)}$ where
$N_{\rm C}$ denotes the number of colours. This allows for a direct
comparison with the semi-numerical results. Figure~\ref{fig:accus}
shows the $\varepsilon$-distributions in absolute normalization
obtained by using double-precision computations for various numbers of
external gluons with alternating polarization states. The phase-space
points are accepted if the generated momenta satisfy the cuts:
$|\eta_n|<2\,,\ p_{\perp,n}>0.1\,|E_1+E_2|$ and $\Delta R_{kn}>0.7$
where $\eta_n$ and $p_{\perp,n}$ respectively denote the
pseudo-rapidity and transverse momentum of the $n$-th outgoing gluon
($n=3,\ldots,N$); the $\Delta R_{kn}$ describe the pairwise geometric
separations in $\eta$\/ and azimuthal-angle space of gluons $k$\/ and
$n$. For $N\ge5$ gluons, a small fraction of events requires higher
than double precision calculations to reliably determine the
master-integral coefficients.\footnote{For $N=6$ and $7$ gluons,
  about 3\% and 10\% of the events need be treated in higher
  precision, respectively.} These phase-space points can be identified
by a simple procedure testing the stability of the solutions for the
bubble coefficients that contribute to the cut-constructible part. As
seen in the plots of Figure~\ref{fig:accus} the gluon one-loop
amplitudes can be determined quite accurately for the bulk of the
events. The tails are sufficiently under control and fall off more
steeply for larger $\varepsilon$-values owing to the veto on points
that yield unstable solutions in double-precision calculations. The
distributions and peak positions of the double poles are rather stable
while those of the single poles and finite parts noticeably shift to
larger $\varepsilon$\/ with an increasing number $N$\/ of gluons. In
all cases the dressed approach is seen to provide more accurate
results than the method relying on the colour decomposition.

\begin{figure}[t!]
\centerline{\psfig{figure=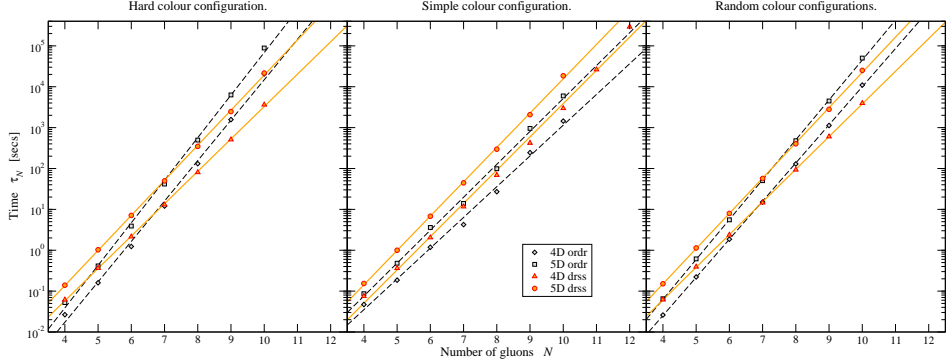,width=53mm,angle=-90}}
\vskip-3mm
\caption{Times $\tau_N$ to compute two $N$-gluon one-loop amplitudes
  for a hard, a simple and random colour configurations using 3.0 and
  2.66 GHz Intel Core2 Duo processors. The solid and dashed curves
  represent fit results for the colour-dressed (drss) and
  colour-ordered (ordr) approach, respectively. The times to compute
  the cut-constructible part of the amplitudes using the 4-dimensional
  (4D) versions of the algorithms are also shown.
  \label{fig:scaling}}
\end{figure}
The complexity of the colour-dressed tree-level recursion relations
and the asymptotic behaviour of the Stirling numbers of the second
kind that govern the growth of the number of the unitarity cuts both
follow an exponential scaling law when more legs are added. The
colour-dressed algorithm is hence expected to scale similarly with
$N$, i.e.\ the computing time $\tau_N$ is proportional to $x^N$ where
$x$\/ is an attribute of the implemented algorithm.
Relying on the colour decomposition, the ordered one-loop amplitude
calculations exhibit polynomial complexity while the number of
orderings that need be evaluated for one colour configuration grows
factorially. The effective growth however will be determined by the
increase in the number of non-vanishing orderings.
Figure~\ref{fig:scaling} depicts how the $N$-gluon one-loop amplitude
computation times scale with increasing $N$\/ for different examples
of assigning the gluons' colour states. Amplitudes with a small number
of gluons, $N\le6$, can always be calculated faster using the colour
decomposition. This remains true as long as there is a sufficient
number of vanishing orderings, i.e.\ one deals with simple colour
configurations only. In all other cases -- for random colour
assignments most importantly -- the colour-dressed method takes over
for $N\ge7$ owing to its milder exponential growth. The curve fitting
for the general case indicates base-values of $x=7.3(1)$ for the
dressed algorithm versus $x=9.5(1)$ when using the colour
decomposition. One notices that the factorial growth caused by the sum
over orderings has been tamed to become effectively exponential.

\begin{figure}[t!]
\psfrag{Yconvergelabel}[b][c][0.45]{
  $\left(\left\langle S^{(0+1)}_{\rm MC}\right\rangle\pm
  \sigma_{\left\langle S^{(0+1)}_{\rm MC}\right\rangle}\right)\;\left\langle\,
  S^{(0+1)}_{\rm col}\right\rangle^{-1}$}
\psfrag{Ylabel}[b][b][0.45]{
  $100\%\ \left.\sigma_{\left\langle S^{(\chi)}_{\rm col/MC}\right\rangle}
  \right/\left\langle S^{(\chi)}_{\rm col/MC}\right\rangle$}
\centerline{
  \psfig{figure=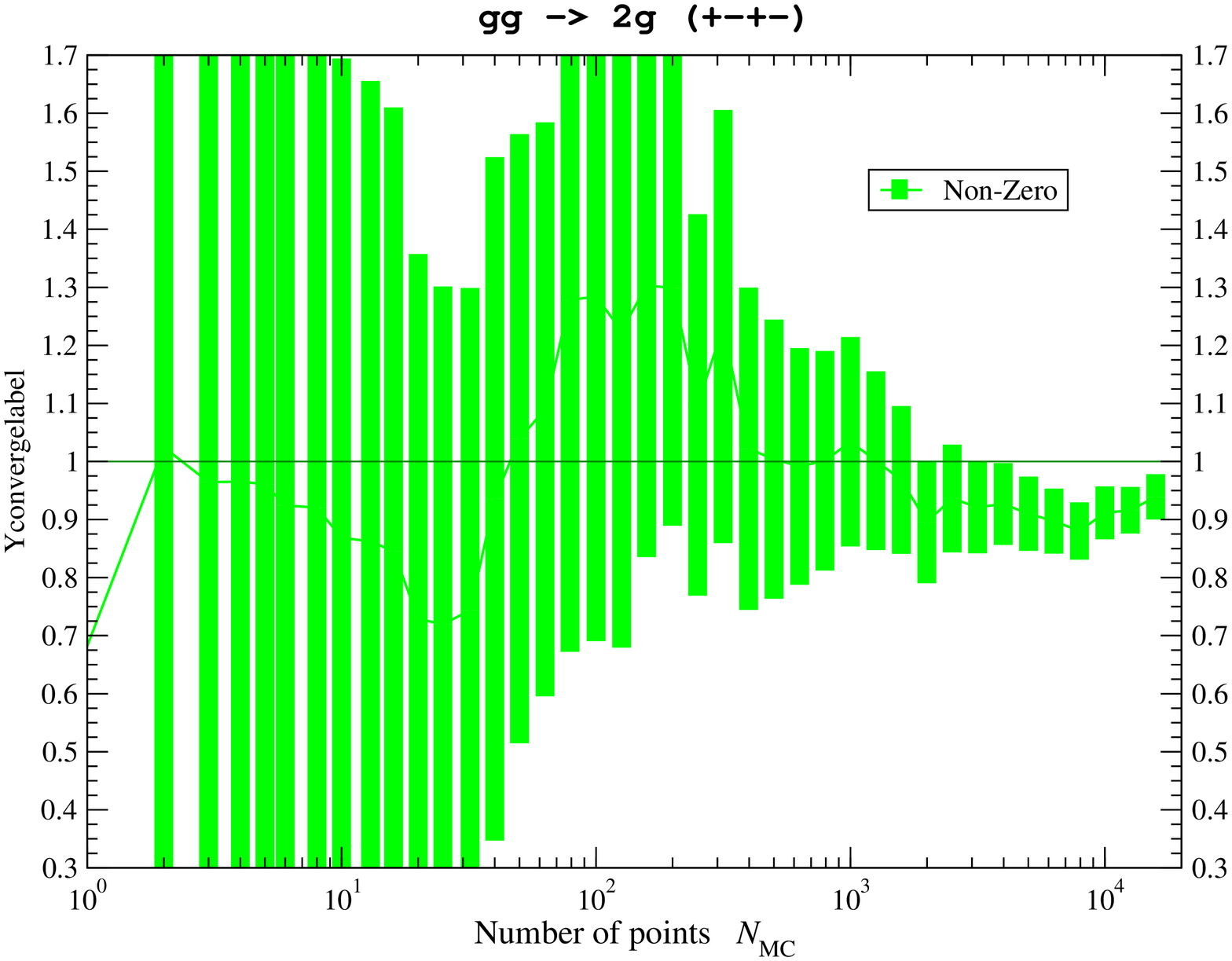,width=52mm,angle=-90}\hskip7mm
  \psfig{figure=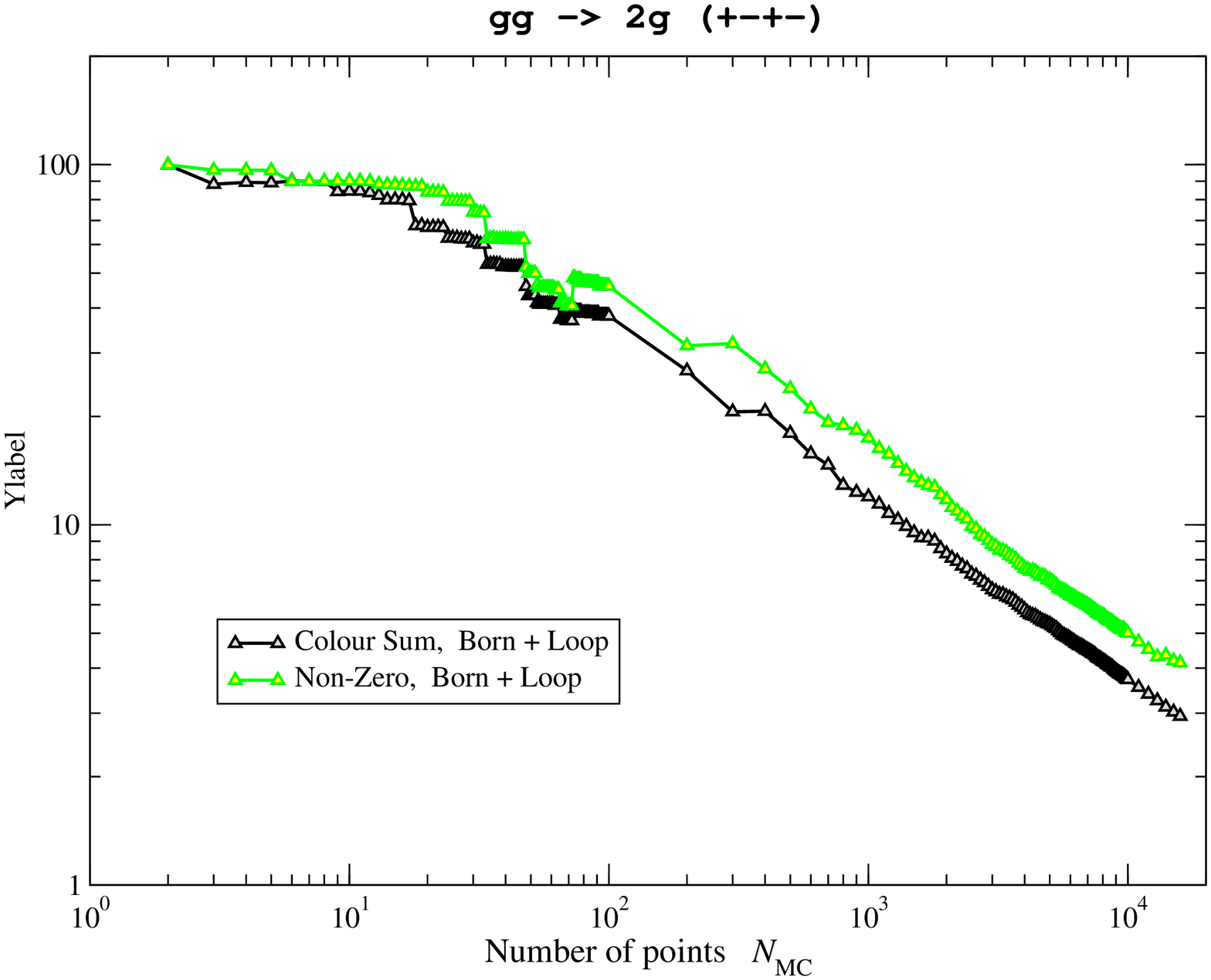,width=52mm,angle=-90}}
\vskip-3mm
\caption{Consistency check for and relative errors of the phase-space
  integrations of colour-summed and colour-sampled amplitudes at the
  Born plus virtual-correction level. The distributions are given as
  functions of the number of generated uniform phase-space points;
  cuts are specified in the text. After 15900 events one obtains
  $\langle S^{(0+1)}_{\rm MC}\rangle/\langle S^{(0+1)}_{\rm col}\rangle
  =0.939\pm0.039({\rm MC})\pm0.028({\rm col})$ with the left plot
  showing errors due to colour sampling only.
  \label{fig:integs}}
\end{figure}
Utilizing the ability of calculating the full one-loop amplitudes for
multiple gluons, one can study the effect of the Monte Carlo sampling
over non-vanishing (``Non-Zero'') colour configurations on the
performance of phase-space integrations for the Born plus virtual
contributions. Figure~\ref{fig:integs} shows the simplest example of
$2\to2$ gluon scattering. The colour-sampled and colour-summed
integrals are defined as
$S^{(0+1)}_\mathrm{MC}=W_\mathrm{col}\times{\cal K}$ and
$S^{(0+1)}_\mathrm{col}=\sum_\mathrm{col}{\cal K}$, respectively,
where the colour weight $W_\mathrm{col}$ depends on the actual colour
configuration (``$\mathrm{col}$'') and the kernel is given by
\be
  {\cal K}\;=\;
  \left|\,{\cal M}^{(0)}\right|^2+\,\frac{\widehat\alpha_s}{2\,\pi}\,
  \Re\left({\cal M}_{\rm fp}^{(1)}\,{{\cal M}^{(0)}}^\dagger\right)
  \qquad\mbox{with}\quad\widehat\alpha_s\equiv0.12\ .
\ee
In the left panel of Figure~\ref{fig:integs} the sampled and summed
integrations are shown to converge after about $10^3$ Monte Carlo
steps, $N_\mathrm{MC}$, while in the right panel their relative errors
are displayed as a function of $N_\mathrm{MC}$. One finds that the
colour sampling only introduces an additional integration uncertainty,
which can be reduced more easily than the full colour sum can be
carried out.

\section*{Acknowledgments}
I thank the organizers for creating a fantastic atmosphere during the
course of the workshop. I also would like to acknowledge my
collaborators Walter Giele and Zoltan Kunszt for many exciting and
important discussions while working on this subject. Fermilab is
operated by the Fermi Research Alliance under CN DE-AC02-07CH11359
with the U.S.\ Department of Energy.


\end{document}